\documentclass[12pt]{article}
\usepackage[utf8]{inputenc}
\usepackage{amsmath}
\usepackage{color}
\renewcommand{\thefootnote}{\fnsymbol{footnote}}
\renewcommand{\thefootnote}{\arabic{footnote}}

\begin{document}
\newcommand{\uRbarI}{\stackrel{\overline{\;\;\;\;}\,_I}{{\bf u}_{_R}}}
\newcommand{\uRbari}{\stackrel{\overline{\;\;\;\;}\,i}{{ u}_{_R}}}
\newcommand{\dRbarI}{\stackrel{\overline{\;\;\;\;}\,_I}{{\bf d}_{_R}}}
\newcommand{\dRbari}{\stackrel{\overline{\;\;\;\;}\,i}{{ d}_{_R}}}
\renewcommand*{\thefootnote}{\fnsymbol{footnote}}

\begin{center}
{\LARGE Physics of Higgs Boson Family \footnote{
		Invited talk at the Conference on New Physics at the LHC, IAS, Singapore, Feb 29 - Mar 4, 2016.} }
\\
\vspace{.3in}
\renewcommand*{\thefootnote}{\arabic{footnote}}
\setcounter{footnote}{0}
Ngee-Pong Chang\footnote{
email: npccc@sci.ccny.cuny.edu}\\

Physics Department, 
City College of CUNY, New York, NY 10031 \footnote{
Permanent Address }
 \\
and  \\
The Graduate Center, CUNY, New York, NY 10016 \\
and \\
Institute of Advanced Studies \footnote{
Visiting Professor } \\
Nanyang Technological Univesity, Singapore

\vspace{.2in}
\end{center}

\begin{abstract}

In the Standard Model, there is the single Higgs field, $\phi$, which gives rise to constituent quark and lepton masses.  The Yukawa coupling is a highly complex set of $3 \times 3$ matrices, resulting in many textures of quark and lepton masses.  \\ 

In this talk, I present a model which transfers the complexity of the Yukawa coupling matrices to a family of Higgs fields, so that the Yukawa coupling itself becomes a simple interaction. \\

In the context of this Enriched Standard Model, we introduce a new $r$-symmetry in the extended $SU(2)_L \times U(1)_Y \times U(1)_R$ model and show how the $125 \;GeV$ and $750 \;GeV$ resonances may be identified with $H$ and $H'$, the key members of the Higgs family, with $H$ being in every way identifed with the SM Higgs. There are interesting consequences of their $2 \gamma$ decay widths. 
\end{abstract}

\begin{flushleft}
\end{flushleft}
\begin{flushleft}
\end{flushleft}

\section{Introduction}

There has been great excitement over the Christmas holidays barely two months ago when CERN 
announced an intriguing hint of a new resonance at $750 \;GeV$.  In my talk today, I would
like to reflect on the implications that such a new resonance at $750 \;GeV$ would have 
for the proposal of the Enriched Standard Model (ESM).\cite{ESM}

In the Standard Model, the quarks acquire mass through the vacuum expectation value of a single $SU(2)$ Higgs doublet, $\phi^{\alpha}$.  The complexity of the texture of quark mass matrices\cite{texture} is attributed to the Yukawa coupling matrix\footnote{
\samepage
Here $i,j$ range over the values $1, 2, 3$.
We have also suppressed, throughout this paper, the $SU(3)_c$ indices on the quarks.  The Higgs fields are as usual taken to be color neutral.
}

\begin{eqnarray}
  {\cal L}_Y &=& - Y^u_{ij} \uRbari \stackrel{\;\;\;\;\;\;j \,\alpha}{q_{_L}} \epsilon_{\alpha \beta} \;\phi^{\beta} + h.c. \nonumber \\
            && - Y^d_{ij} \dRbari \stackrel{\;\;\;\;\;\;j \,\alpha}{q_{_L}} \phi_{\alpha} + h.c.
\end{eqnarray}
where we have introduced for later convenience the convention
\begin{eqnarray}
		\phi_{\alpha}	&\equiv&  \left( \phi^{\alpha} \right)^{\dagger}
\end{eqnarray}

\noindent Much of the Standard Model phenomenology is dedicated to the determination of the magnitudes and phases of the $CKM$ matrix that can be derived from the $u$, $d$ Yukawa coupling matrices. In particular, there are many unexplained hierarchies in the quark masses, and the Wolfenstein hierarchies in the associated $CKM$ matrices. \\

\noindent In what I call the ESM, we transfer the complexity of the Yukawa coupling to the $SU(3)$ family of Higgs fields, so that the Yukawa interaction is now
\begin{eqnarray}\left.\begin{array}{rcl}
  {\cal L}_Y &=& - h_u \;\uRbari \; \stackrel{\;\;\;\;\;\;j \,\alpha}{q_{_L}} \epsilon_{\alpha \beta} \;\phi^{\beta}_{_{ij}} + h.c.  \\
            && - h_d \;\dRbari \; \stackrel{\;\;\;\;\;\;j \,\alpha}{q_{_L}}  \;\widehat{\phi}\,{}_{ij \,\alpha} + h.c.
            \end{array} \right\} \label{Yukawa u-d int}
\end{eqnarray}
Note that in so doing we have also added the distinction between the Higgs fields, $\widehat{\phi}\,{}_{ij \,\alpha}$, coupled to the down-quark family versus the original  $\phi^{\alpha}_{ij}$ associated with the up-quark family.  From a certain point of view, it looks counter-intuitive to double down on the number of Higgs fields when phenomenologically there was such a great difficulty in even finding one Higgs.  The pay-off comes when you realize there is a greater symmetry that results from so doing.

\section{r-Symmetry}

To make the model more attractive, we make the simple requirement 
\begin{eqnarray}
		h_u	&=& h_d
\end{eqnarray} 
so that
\begin{eqnarray}
{\cal L}_Y &=& -h_q \left( \stackrel{\overline{\;\;\;\,}\,_i}{ u_{_R}} \;q_{L}^{j \,\alpha} \;\phi^{\beta}_{_{ij}} 
					\;\epsilon_{\alpha \beta}   \;+\; \stackrel{\overline{\;\;\;\,}\,_i}{ d_{_R}} \;q_{L}^{j \,\alpha} 
					\; \widehat{\phi}\,{}_{_{ij}\alpha} \;\;\right) \;\;\;\;+ \; h.c. 
\end{eqnarray}

\noindent This requirement may look odd, as everyone knows that the down quark in each family is much lighter than the up quark.  But, for this Enriched Standard Model, with the two families of Higgs fields, $\phi^{\alpha}$, and $\widehat{\phi}\,{}_{\alpha}$, the difference in the physical up and down quark masses may be attributed to the difference in vacuum expectation values of the corresponding Higgs fields. \\

\noindent To implement this requirement we extend the enriched Standard Model to the gauge group $SU(2)_L \times U(1)_Y \times U(1)_R$ and impose a new $r$-symmetry on the full Lagrangian.   \\

\begin{center}
\begin{tabular}{|c|c|c|c|| c| c|c| c |}
		\hline
			& $Y_R$	&	$Y'$ & $(I_{3})_{L}$   & & $Y_R$	&	$Y'$ & $(I_{3})_{L}$  \\  \hline
		$\phi\,{}^{+}_{ij}$	&	$+1/2$	&	& $+ 1/2$ 	 & $u^{j}_{L}$ &  & $+1/6$ & $+1/2$ \\
		$\phi\,{}^{o}_{ij}$	&	$+ 1/2$ & & $-1/2$ 		 & $d^{j}_{L}$ &  & $+1/6$ & $-1/2$\\
		$\widehat{\phi}\,{}^{+ \,ij}$	& $+ 1/2$ 	 &  & $+ 1/2$ & $u_{_{Ri}}$ & $+1/2$  & $+1/6$ &  \\
		$\widehat{\phi}\,{}^{o \,ij}$  &	$+ 1/2$		 &  & $- 1/2$ & $d_{_{Ri}}$ & $-1/2$  & $+1/6$ & \\
		\hline
\end{tabular}
\end{center}

\noindent The covariant derivatives for the quark \& Higgs fields now read as\footnote{ 
		Here $a,b = 1,2$ refer to the flavor of the right-handed quark families $q^{a}_{_R \,i}$, while $\alpha = 1,2$ refer as usual to 
		the flavor of the left-handed $q^{j}_{L}$ quark families, and  $i,j=1,2,3$ refer to the families. 
}
\begin{eqnarray}
	D_{\mu} \;q^{j \,\alpha}_L	&=& \partial_{\mu} \;q^{j \,\alpha}_{L} + i \;\frac{g}{2} \;\left( \vec{\tau} \cdot \vec{W} \right)^{\alpha}_{\beta} \;q^{j \,\beta}_{L} + i \;\frac{g'}{6} \;B^{'}_{\mu} \;q^{j \,\alpha}_{L}  \\
	D_{\mu} \;q^{a}_{R\,i}	&=& \partial_{\mu} \;q^{a}_{R\,i} + i \;\frac{g_R}{2} \;\Bigl( \tau_3 \;B_{R \,\mu} \Bigr)^{a}_{b} \;q^{b}_{R\,i} + i \;\frac{g'}{6} \;B^{'}_{\mu} \;q^{a}_{R\,i} \\
	D_{\mu} \,\phi^{\alpha}_{ij} &=& \partial_{\mu}\,\phi^{\alpha}_{ij} \,+\, i \,\frac{g}{2} \,\left( \vec{\tau} \cdot \vec{W} \right)^{\alpha}_{\beta} \,\phi^{\beta}_{ij}  \,+\, i \,\frac{g_R}{2} \;B_{R\mu} \,\phi^{\alpha}_{ij}  \\
	D_{\mu} \,\widehat{\phi}\,{}_{\alpha \,ij} &=& \partial_{\mu}\,\widehat{\phi}\,{}_{\alpha \,ij} \,-\, i \,\frac{g}{2} \,\left( \vec{\tau} \cdot \vec{W} \right)_{\alpha}^{\beta} \,\widehat{\phi}\,{}_{\beta \,ij}  \,-\, i \,\frac{g_R}{2} \;B_{R\mu} \,\widehat{\phi}\,{}_{\alpha \,ij} \label{covar deriv I}
\end{eqnarray}

\noindent The full Lagrangian is invariant under the $r$-symmetry

\begin{eqnarray}
\left. 
\begin{array}{rclcrcl}
	u_{_{Ri}}		& \rightarrow &  \;\;\;\;\;\; d_{_{Ri}}  & & 	d_{_{Ri}}		& \rightarrow & - u_{_{Ri}} \\
	\phi\,{}^{\alpha}_{ij} & \rightarrow & - \epsilon^{\alpha \beta} \;\widehat{\phi}\,{}_{\beta \,ij} & &
	\widehat{\phi}\,{}_{\alpha \,ij} & \rightarrow & - \epsilon_{\alpha \beta} \;\phi\,{}^{\beta}_{ij}
\end{array}
	\right\} &&
\end{eqnarray}
and for the gauge fields
\begin{eqnarray}
\left. \begin{array}{ccl}
	B_R \,{}_{\mu} &\rightarrow& - \; B_R \,{}_{\mu}  \\
	\vec{W}_{\mu}	&\rightarrow& + \;\vec{W}_{\mu} \\
	B^{'}_{\mu}	&\rightarrow&  + \;B^{'}_{\mu}	
	\end{array} \right\} &&
\end{eqnarray}

\noindent This extension of the Standard Model parallels the $SU(2)_{L} \times SU(2)_R \times U(1)_Y$ of Mohapatra \& Senjanovic\cite{Mohapatra}.  Instead of the full set of $SU(2)_R$ gauge bosons, however, we have only the neutral $B_R$ gauge boson.
To have the correct neutrino phenomenology, we introduce a set of very heavy Higgs fields, $ \Delta^{R \,,ij}_{ab}$.

\section{Include Leptons}

\noindent To include leptons, we introduce the compact notation for the Higgs fields
\begin{eqnarray}
\phi^{\beta}_{a \,,ij} &=& \left(  
		\begin{array}{rcc}
	\widehat{\phi}\,{}_{(o) \,ij} && \phi^{(+)}_{ij}   \\
	- \widehat{\phi}\,{}_{(-) \,ij} &&  \phi^{(o)}_{ij} 
		\end{array}  \right)^{\beta}_{a} 
\end{eqnarray}
using the convention that
\begin{eqnarray}
\left.  	\begin{array}{lcl}
	\phi^{ij}_{\alpha}  &\equiv&  \left( \phi^{\alpha}_{ij} \right)^{\dagger}  \\ 
	\widehat{\phi}\,{}^{\alpha \,ij}  &\equiv&  \left( \widehat{\phi}_{\alpha \,ij} \right)^{\dagger}  
					\end{array} \right\}
\end{eqnarray}

\noindent Here in addition to the Dirac mass terms for the leptons, we introduce the Majorana mass terms for the leptons.  The complete fermion Yukawa Lagrangian now reads
\begin{eqnarray}
\left.  	\begin{array}{rcl}
	{\cal L}_{Y} &=& \;-\; h_{q} \; \stackrel{\overline{\;\;\;\;\;}\;\;_i}{{ q}_{_{R \,a}}} \;q^{j \alpha}_{L} \;\epsilon_{\alpha \beta} \;\phi^{\beta}_{b \,ij} \;\epsilon^{ab} \;+\; h.c.  \\
			& & \;-\; h_{\ell} \; \stackrel{\overline{\;\;\;\;}\;_i}{{ \ell}_{R \,a}} \;\ell^{j \alpha}_{L} \;\epsilon_{\alpha \beta} \;\phi^{\beta}_{b \,ij} \;\epsilon^{ab} \;+\; h.c.  \\
			& & \;+\; \frac{1}{2}  \;\widetilde{\ell}^{\,a}_{R \,i} \;{\cal C} \;\ell^{\,b}_{R \,j} \;\Delta^{R \,,ij}_{ab} \;\;+\; h.c.  
					\end{array} \right\}
\end{eqnarray}
Under $r$-symmetry, the fermion fields transform as ( $\left( i \sigma_2 \right)^a_b \;=\; \epsilon_{ab}$ )
\begin{eqnarray}
\left.  	\begin{array}{rcl}
		q^{a}_{_{R \,i}}		& \rightarrow &  \left( i \sigma_2 \right)^{a}_{b} \;q^{b}_{_{R \,i}} \\
		q^{j \,\alpha}_{L}	& \rightarrow &  q^{j \,\alpha}_{L} \\
		\ell^{a}_{R \, i} &\rightarrow& \left( i \sigma_2 \right)^{a}_{b} \;\ell^{b}_{R \,i} \\
		\ell^{j \,\alpha}_{L} & \rightarrow & \ell^{j \,\alpha}_{L}
					\end{array} \right\}
\end{eqnarray}
while the Higgs fields transform as
\begin{eqnarray}
\left.  	\begin{array}{rcl}
		\phi^{\beta}_{b \,ij}	&\rightarrow&  \left( - i\sigma_2 \right)^{c}_{b} \;\phi^{\beta}_{c \,ij}  \\
		\Delta^{R \,ij}_{ab} &\rightarrow& \left( i \sigma_2 \right)^{c}_{a} \;\Delta^{R \,ij}_{cd} 
						\;\left( i \sigma_2 \right)^{d}_{b} 
					\end{array} \right\}
\end{eqnarray}
or, specifically for the new $\Delta^{R \,,ij}_{ab}$ fields under $r$-symmetry
\begin{eqnarray}
\left.  	\begin{array}{rcl}
		\Delta^{R \,ij}_{11}  & \longleftrightarrow & \;\;\Delta^{R \,ij}_{22}  \\
		\Delta^{R \,ij}_{12}  & \longleftrightarrow & - \Delta^{R \,ij}_{21} 
					\end{array} \right\}
\end{eqnarray}

\noindent The covariant derivatives of the Higgs fields now read in totality
\begin{eqnarray}
	\left.  \begin{array}{lcl}
	D_{\mu} \phi^{\beta}_{ij \,a} &=&  \partial_{\mu} \phi^{\beta}_{ij \,a} + i \displaystyle \frac{g}{2} \left( \vec{\tau} \cdot \vec{W}_{\mu} \right)^{\beta}_{\alpha} \phi^{\alpha}_{ij \,a}  - i \frac{g_{_R}}{2} \left( \sigma_3 \right)^{a'}_{a} B_{_R \,\mu} \phi^{\alpha}_{ij \,a'}      \\
	D_{\mu} \Delta^{ij}_{R \,ab } &=&  \partial_{\mu} \Delta^{ij}_{R \,ab } - i \displaystyle \frac{g_{_R}}{2} \left( \sigma_3 \right)^{a'}_{a} B_{_R \,\mu}\Delta^{ij}_{R \,a'b }  \\
		&&  - i \displaystyle \frac{g_{_R}}{2} \left( \sigma_3 \right)^{b'}_{b} B_{_R \,\mu} \Delta^{ij}_{R \,ab' } 
				+ i g' B_{\mu} \Delta^{ij}_{R \,ab }    
				\end{array}  \right\} && \label{covar deriv II}
\end{eqnarray}

\noindent Following Mohapatra \& Senjanovic\cite{Mohapatra}, we implement the symmetry breaking pattern such that the vacuum expectation values of the different Higgs fields 
\begin{eqnarray}
	< \Delta^{ij}_{R \,ab} > &=& \left(  
			\begin{array}{ccc}
			\upsilon_R^{ij} && 0   \\
			0 &&  0
			\end{array}  \right)_{ab} \\
	 < \phi^{\beta}_{b \,ij} > &=& \left(
			\begin{array}{ccc}
			\widehat{\upsilon}\,{}_{ij} & 0 \\
			0 & \upsilon_{ij} 
			\end{array} \right)^{\beta}_{b}
\end{eqnarray}
possess the hierarchy
\begin{eqnarray}
\upsilon_{R}  \;>> \; \upsilon \;>>\; \widehat{\upsilon} 
\end{eqnarray}
where we have used the simplified notation
\begin{eqnarray}
\left.  \begin{array}{lcl}
		\upsilon^2	&=& \upsilon^{ij} \;\upsilon_{ij}  \\
		\widehat{\upsilon}\,{}^2	&=& \widehat{\upsilon}\,{}_{ij} \;\widehat{\upsilon}\,{}^{ij}  \\
		\upsilon^2_{R}	&=& \upsilon^{ij}_{_R} \;\upsilon_{ij \,_R}
				\end{array} \right\}  \label{vev}
\end{eqnarray}

\section{Gauge Bosons Acquire Mass}
\noindent From the covariant derivatives in eq.(\ref{covar deriv II}), we arrive at the physical gauge bosons ( $ \upsilon^2_{t} \;\equiv\; \upsilon^2 + \widehat{\upsilon}\,{}^2$ )

\begin{eqnarray}
	\left.   \begin{array}{lcl}
	Z_{\mu} &=& \cos{\theta_W} \;W^{3}_{\mu}  \;-\; \sin{\theta_W} \; \displaystyle\frac{\left( g' \,B_{R \mu} + g_{_R} \,B_{\mu} \right)  }{ \sqrt{g_{_R}^2 + g'\,{}^2} } \;+\; O \Bigl( \frac{\upsilon_{t}^2}{\upsilon_R^2} \Bigr)  \\
	A_{\mu} &=& \sin{\theta_W} \;W^{3}_{\mu} \;+\; \cos{\theta_W} \; \displaystyle \frac{\left( g' \,B_{R \mu} + g_{_R} \,B_{\mu} \right) } {\sqrt{g_{_R}^2+ g'\,{}^2}}  \\
	Z_{R \mu} &=& \displaystyle \frac{g_{_R} \,B_{R \mu} - g' \,B_{\mu}}{\sqrt{g_{_R}^2 + g'\,{}^2}}  \;+\; O \Bigl( \frac{\upsilon_{t}^2}{\upsilon_R^2} \Bigr)  
							\end{array} \right\} && \label{gauge boson}
\end{eqnarray}
with the masses given by
\begin{eqnarray}
	m_W^2 &=& \frac{1}{2} \,g^2 \upsilon_t^2    \\
	m_Z^2 &=& \displaystyle \frac{m_W^2}{\cos^2 \theta_W }   \\
		&=& \frac{\upsilon^2_t}{2} \; \left( g^2 + g\,{}^{\prime \,2}_s \right)  \label{m_Z std} \\
	m_{Z_R}^2 &=& 2 \left( \upsilon_R^2  \right) \;\left( g_{_R}^2 + g\,{}^{\prime \,2}_s \right) \;+\; O(\upsilon^2) 
\end{eqnarray}
where
\begin{eqnarray}
	g\,{}^{\prime}_s	&=&  \displaystyle \frac{ g_{_R} \,g\,{}^{\prime}}{ \sqrt{ g^2_{_R} + g\,{}^{\prime \,2}_s   } } 
\end{eqnarray}
or
\begin{eqnarray}
	\displaystyle \frac{1}{g\,{}^{\prime \,2}_s }  &=& \frac{1}{g^2_{_R} } + \frac{1}{ g\,{}^{\prime \,2} }  \label{g's relation}
\end{eqnarray}
From eq.(\ref{m_Z std}) we see that the $g'_{s}$ is actually the $U(1)_Y$ coupling of the Standard Model, and eq.(\ref{g's relation}) gives its relationship with the couplings of the extended $SU(2)_L \times U(1)_Y \times U(1)_R$ model.  \\

\noindent These relationships are a manifestation of the decoupling theorem of Georgi-Weinberg\cite{Georgi-Weinberg}.  In eq.(\ref{gauge boson}) we see how in the limit of $ \upsilon_{t} / \upsilon_R \rightarrow 0$, 
\begin{eqnarray}
		\displaystyle \frac{ g' B_{R \mu} + g_{_R} B_{\mu} }{ \sqrt{ g^2_{_R} + g'^2 }} &\rightarrow&  B_{Y\mu}
\end{eqnarray}
where $B_{Y \mu}$ is the $U(1)_Y$ gauge field of the usual $SU(2)_L \times U(1)_Y$ group.

\section{A Simple Higgs Potential}

As noted already by Mohapatra and Senjanovic\cite{Mohapatra}, neutrino phenomenology requires that the Higgs fields $\Delta^{ij}_{R \,ab}$ be associated with a mass scale that is much higher than the Higgs $\phi^{\beta}_{a \,ij}$.  For our purposes, the Georgi-Weinberg decoupling theorem\cite{Georgi-Weinberg} enables us to focus on the low energy phenomenology associated with the $\phi^{\beta}_{a \,ij}$.  \\

\noindent Rather than work with a most general for the Higgs potential, We turn to a particularly simple form for the Higgs potential.  \\

\noindent We consider the potential
\begin{eqnarray}
		V &=& V_{\phi} + V_{\Delta}    \label{Higgs Full Pot}
\end{eqnarray}
where $V_{\phi}$ involves the lighter $\phi^{\alpha}_{ij}$ and $\widehat{\phi}\,^{\alpha \,k\ell}$ fields, while $V_{\Delta}$ involves the heavy $\Delta^{ij}_{R \,ab}$ fields.  
For the general non-degenerate case, we introduce the three coupling constants, $\lambda_1, \lambda_2, \lambda_3$ in the maximally symmetric potential
\begin{eqnarray}
		V_{\phi}	&=& + \displaystyle \frac{\lambda_1}{2} \left( \epsilon_{\alpha \beta} \;\phi\,{}^{\alpha}_{a, \,ij} \;\phi\,{}^{\beta}_{b, \,k\ell}  \right) 
		\left( \epsilon^{\alpha' \beta'} \;\phi\,{}_{\alpha'}^{a, \,ij} \;\phi\,{}_{\beta'}^{b, \,k\ell}  \right)  \nonumber \\
							& & + \displaystyle \frac{\lambda_2}{2} \left( \phi\,{}^{\alpha}_{a, \,ij} \;\phi\,{}^{b, \,k\ell}_{\alpha} \right) 
		\left( \phi\,{}^{\beta}_{b, \,k\ell} \;\phi\,{}^{a, \,ij}_{\beta} \right) - \displaystyle \frac{\lambda_3}{4} \left( \phi\,{}^{\alpha}_{a, \,ij} \;\phi\,{}^{a, \,ij}_{\alpha} \right)^2   
\end{eqnarray}
The symmetry is broken through the vacuum expectation values of the $\phi\,{}^{\beta}_{a, \,ij}$ fields, as given in eq.(\ref{vev}) above.  \\

\noindent The Higgs potential $V_{\phi}$ around the new vacuum takes the form
\begin{eqnarray}
		V_{\phi}	&=& + \displaystyle \frac{\lambda_1}{2} \left( \epsilon_{\alpha \beta} 
										\left[ \;\phi\,{}^{\alpha}_{a, \,ij} \;\phi\,{}^{\beta}_{b, \,k\ell} 
												- \upsilon\,{}^{\alpha}_{a, \,ij} \;\upsilon\,{}^{\beta}_{b, \,k\ell} \right] \right) \times \nonumber \\
							& &   \;\;\;\;\;\;\; \left( \epsilon^{\alpha' \beta'} \left[ \;\phi\,{}_{\alpha'}^{a, \,ij} \;\phi\,{}_{\beta'}^{b, \,k\ell} 
												- \upsilon\,{}_{\alpha}^{a, \,ij} \;\upsilon\,{}_{\beta}^{b, \,k\ell} \right]  \right)  \nonumber \\
							& & + \displaystyle \frac{\lambda_2}{2} 
										\left( \left[ \phi\,{}^{\alpha}_{a, \,ij} \;\phi\,{}^{b, \,k\ell}_{\alpha}  
												- \upsilon\,{}^{\alpha}_{a, \,ij} \;\upsilon\,{}^{b, \,k\ell} \right] \right) \times  \nonumber \\
							& &   \;\;\;\;\;\;\; \left( \left[ \phi\,{}^{\beta}_{b, \,k\ell} \;\phi\,{}^{a, \,ij}_{\beta} 
												- \upsilon\,{}^{\beta}_{b, \,k\ell} \;\upsilon^{a, \,ij}_{\beta} \right]  \right) \nonumber \\
							& & - \displaystyle \frac{\lambda_3}{4} \Bigl( \phi\,{}^{\alpha}_{a, \,ij} \;\phi\,{}^{a, \,ij}_{\alpha} 
												- \upsilon^2 - \widehat{\upsilon}\,{}^2 \Bigr)^2   \label{Higgs Pot}
\end{eqnarray}

\noindent Likewise, the Higgs potential involving the heavy fields takes the form
\begin{eqnarray}
\left.		\begin{array}{lcl}
			V_{\Delta} &=& +  \;\lambda_4 \;\left( \Delta^{ij}_{R \,ab} \,\Delta^{R \,ab}_{ij} \;-\; \upsilon^2_{R}  \right)^2  \\
					& & +  \;\lambda_5 \;\left( \Delta^{ij}_{R \,ab} \,\Delta^{k\ell}_{R \,cd} \;\epsilon^{ac} 
									\epsilon^{bd} \right)  
									\,\left( \Delta_{ij}^{R \,a'b'} \,\Delta_{k\ell}^{R \,c'd'} \;\epsilon_{a'c'} \epsilon_{b'd'} \right)\\
					& & +  \;\lambda_6 \;\upsilon^2_{R} \left[  \Delta^{ij}_{R \,ab} \left( \sigma_1 \right)^{a}_{a'}  \Delta^{R \,a'b'}_{ij} 
					\left( \sigma_1 \right)^{b}_{b'}  \right. \\
					& &\left. \;\;\;\;\;\;\;\;\;\;\; -  \Delta^{ij}_{R \,ab} \left( i \sigma_2 \right)^{a}_{a'}  \Delta^{R \,a'b'}_{ij} 
						\left( i \sigma_2 \right)^{b}_{b'}   \right]
				\end{array} \right\}
				\label{Higgs Pot_'}
\end{eqnarray}

\noindent  By construction, this Higgs potential is stable about the broken vacuum with $\upsilon_{ij}$ and $\widehat{\upsilon}\,{}_{k\ell}$.

\section{Quark Mass Diagonal (QMD) Basis}

To work out the mass spectrum for the Higgs potential of eq.(\ref{Higgs Pot}), we need to go to the basis where the 
physical quark fields have diagonal mass matrices, where
\begin{eqnarray}
	\stackrel{\overline{\;\;\;\;}\,_i}{{ u}_{_R}} &=& \stackrel{\overline{\;\;\;\;}\,_I}{{\bf u}_{_R}}\;\left( V^{*}_{uR} \right)^i_{_I} \nonumber \\
	u^j_{_L} &=& \left( V_{uL} \right)^j_{_J} \;{\bf u}^{_J}_{_L} \nonumber \\
	\stackrel{\overline{\;\;\;\;}\,i}{{ d}_{_R}} &=& \stackrel{\overline{\;\;\;\;}\,_I}{{\bf d}_{_R}}\;\left( V^{*}_{dR} \right)^i_{_I} \nonumber \\
	d^j_{_L} &=& \left( V_{dL} \right)^j_{_J} \;{\bf d}{}^{_J}_{_L} \nonumber 
\end{eqnarray}
For simplicity, we make the assumption
\begin{eqnarray}
		V_{uL}	&=& V_{uR}  \\
		V_{dL}	&=& V_{dR}
\end{eqnarray}
so that the CKM matrix takes the form
\begin{eqnarray}
 V^{i}_{j} &=& \Bigl( \left( V_u \right)^{\dagger} \;V_d \Bigr)^{i}_{j} 
					\label{CKM defined}
\end{eqnarray}
In this QMD basis, the Higgs fields take the form
\begin{eqnarray}
	\phi^{(o)}_{ij} &=& \left(\widetilde{V}_u \right)^{i'}_{i} \;\Phi^{(o)}_{i'j\,'}   \;\left( V^{\dagger}_u \right)^{j\,'}_{j}     \\
	\phi^{(+)}_{ij} &=& \left(\widetilde{V}_u \right)^{i'}_{i} \;\Phi^{(+)}_{i'j\,'}   \;\left( V^{\dagger}_d \right)^{j\,'}_{j}     
\end{eqnarray}

\begin{eqnarray}
	\widehat{\phi}\,{}^{(o) \,ij} &=& \bigl({V}^{*}_d \bigr)^{i}_{i'} \;\widehat{\Phi}\,{}^{(o)\,i'j\,'}   \;\bigl( V_d \bigr)^{j}_{j\,'}     \\
	\widehat{\phi}\,{}^{(+)\,ij} &=& \bigl( V^{*}_d \bigr)^{i}_{i'} \;\widehat{\Phi}\,{}^{(+)\,i'j\,'}   \;\bigl( V_u \bigr)^{j}_{j\,'}     
\end{eqnarray}
and the $\Phi_{ij}, \widehat{\Phi}\,{}^{ij}$ fields have vacuum expectation values $w_{ij}, \,\widehat{w}\,{}^{ij}$ that are diagonal, with
\begin{eqnarray}
	< \Phi^{(o)}_{ij} > &=&  w_{ij} \;= \left(  \begin{array}{ccc}
				\upsilon_1 & 0 & 0 \\
				0 & \upsilon_2 & 0 \\
				0 & 0 & \upsilon_3 
				\end{array} \right)_{ij} 
\end{eqnarray}
and 
\begin{eqnarray}
	< \widehat{\Phi}\,{}^{(o) \,ij} > &=&  \widehat{w}\,{}^{ij} = \left(  \begin{array}{ccc}
				\widehat{\upsilon}_1 & 0 & 0 \\
				0 & \widehat{\upsilon}_2 & 0 \\
				0 & 0 & \widehat{\upsilon}_3 
				\end{array} \right)^{ij} 
\end{eqnarray}

\noindent The Higgs potential in eq.(\ref{Higgs Pot}) gives rise to a rich mass spectrum.  It involves the full complexity of the texture of the vacuum expectation values, and poses a daunting task for the timid explorer.  In my earlier works, I had explored the spectrum in the leading hierarchy by simply setting $\upsilon_{33} = \upsilon, \widehat{\upsilon}\,{}_{33} = \widehat{\upsilon}$, and letting all the other vacuum expectation values be zero.  \\

In this talk, I will set forth a more complete analysis of the mass spectrum for the general mass hierarchy.  \\

\section{ Neutral Higgs Boson Spectrum}

For the neutral Higgs sector, we express the resulting spectrum in terms of the hermitian fields, $h_{ij}$ and $z_{ij}$, with		
\begin{eqnarray}
		\left.   \begin{array}{lcl}
		\Phi^{o}_{ij}	&=&  \displaystyle \frac{h_{ij} - i \;z_{ij}}{\sqrt{2}}   \\
		\widehat{\Phi}\,{}^{o \,ij}	&=& \displaystyle \frac{\widehat{h}\,{}_{ij} - i \;\widehat{z}\,{}_{ij}}{\sqrt{2}}  
						\end{array}  \right\}
\end{eqnarray}

The flavor-diagonal family of neutral scalar Higgs bosons may be expressed in terms of the orthonormal basis $h$ eigenstates
\begin{eqnarray}
		\left.   \begin{array}{lcl}
	h_a	&=& \left( \;\;\upsilon_3 \,h_{33} \;+ \;\;\;\;\;\; \left( \upsilon_2 \,h_{22} +  \upsilon_1 \,h_{11} \right) \right) / \upsilon   \\
	h_b	&=& \bigl( - \upsilon_{\perp} \,h_{33} + \displaystyle \frac{\upsilon_3}{\upsilon_{\perp}} 
			\left( \upsilon_2 \,h_{22} + \upsilon_1 \,h_{11} \bigr) \right) / \upsilon   \\
	h_c	&=&  \;\;\;\;\;\;\;\;\;\;\;\;\;\;\;\;\;\;\;\;\;\;\;\left( - \upsilon_1 \,h_{22} + \upsilon_2 \,h_{11}  \right) / \upsilon_{\perp}   \vspace{.1in} \\
	\widehat{h}\,{}_a	&=& \left( \;\; \widehat{\upsilon}\,{}_3 \widehat{h}\,{}_{33} 
							+ \;\;\;\;\;\;\; \left(\widehat{\upsilon}\,{}_2 \widehat{h}\,{}_{22} 
							+  \widehat{\upsilon}\,{}_1 \widehat{h}\,{}_{11} \right) \right) / \widehat{\upsilon}{}   \\
	\widehat{h}\,{}_b	&=& \bigl( - \widehat{\upsilon}\,{}_{\perp} \widehat{h}\,{}_{33} 
							+ \displaystyle \frac{ \widehat{\upsilon}\,{}_3 }{\widehat{\upsilon}\,{}_{\perp}} \left( \widehat{\upsilon}\,{}_2 \widehat{h}\,{}_{22} 
							+ \widehat{\upsilon}\,{}_1 \widehat{h}\,{}_{11} \right) \bigr) / \widehat{\upsilon}{}
			 \\
	\widehat{h}\,{}_c	&=&  \;\;\;\;\;\;\;\;\;\;\;\;\;\;\;\;\;\;\;\;\;\;\;\;\left( - \widehat{\upsilon}{}_1 \,\widehat{h}_{22} 
	        + \widehat{\upsilon}{}_2 \,\widehat{h}_{11}  \right) / \widehat{\upsilon}{}_{\perp}    
						\end{array}  \right\}
\end{eqnarray}
where
\begin{eqnarray}
		\left. \begin{array}{lcr}
		\upsilon	&=& \sqrt{ \upsilon^2_3 + \upsilon^2_2 + \upsilon^2_1 }  \\
		\upsilon_{\perp} &=& \sqrt{ \upsilon^2_2 + \upsilon^2_1 } \\
		\widehat{\upsilon} &=& \sqrt{ \widehat{\upsilon}^2_3 + \widehat{\upsilon}{}^2_2 + \widehat{\upsilon}{}^2_1 }  \\
		\widehat{\upsilon}{}_{\perp} &=& \sqrt{ \widehat{\upsilon}{}^2_2 + \widehat{\upsilon}{}^2_1 } 
					\end{array} \right\}
\end{eqnarray}
Likewise, the flavor-diagonal pseudoscalar Higgs bosons may be expressed in terms of the orthonormal basis $z$ eigenstates
\begin{eqnarray}
		\left.   \begin{array}{lcl}
	z_a	&=& \left( \;\; \upsilon_3 \,z_{33} \;+ \;\;\;\;\;\;\left( \upsilon_2 \,z_{22} +  \upsilon_1 \,z_{11} \right) \right) / \upsilon   \\
	z_b	&=& \bigl( -\upsilon_{\perp} \,z_{33} + \displaystyle \frac{\upsilon_3}{\upsilon_{\perp}} 
			\left( \upsilon_2 \,z_{22} + \upsilon_1 \,z_{11} \bigr) \right) / \upsilon   \\
	z_c	&=&  \;\;\;\;\;\;\;\;\;\;\;\;\;\;\;\;\;\;\;\;\;\;\;\;\left( - \upsilon_1 \,z_{22} + \upsilon_2 \,z_{11}  \right) / \upsilon_{\perp}   \vspace{.1in} \\
	\widehat{z}\,{}_a	&=& \left( \;\;\widehat{\upsilon}\,{}_3 \widehat{z}\,{}_{33} 
							\;+ \;\;\;\;\;\; \left(\widehat{\upsilon}\,{}_2 \widehat{z}\,{}_{22} 
							+  \widehat{\upsilon}\,{}_1 \widehat{z}\,{}_{11} \right) \right) / \widehat{\upsilon}{}   \\
	\widehat{z}\,{}_b	&=& \bigl( - \widehat{\upsilon}\,{}_{\perp} \widehat{z}\,{}_{33} 
							+ \displaystyle \frac{ \widehat{\upsilon}\,{}_3 }{\widehat{\upsilon}\,{}_{\perp}} \left( \widehat{\upsilon}\,{}_2 \widehat{z}\,{}_{22} 
							+ \widehat{\upsilon}\,{}_1 \widehat{z}\,{}_{11} \right) \bigr) / \widehat{\upsilon}{}
			 \\
	\widehat{z}\,{}_c	&=&  \;\;\;\;\;\;\;\;\;\;\;\;\;\;\;\;\;\;\;\;\;\;\;\;\left( - \widehat{\upsilon}{}_1 \,\widehat{z}_{22} 
	        + \widehat{\upsilon}{}_2 \,\widehat{z}_{11}  \right) / \widehat{\upsilon}{}_{\perp}    
						\end{array}  \right\}
\end{eqnarray}

In terms of these basis states, the spectrum of neutral Higgs mass eigenstates is given by\footnote{\samepage
Here I have adopted the notation as employed by the $2HDM$ and $MSSM$ literature.  The masses for the Goldstone bosons are given in the 't Hooft gauge, $(\xi=1)$ .
}

\begin{center}
Table I \;\;\;\;
\begin{tabular}{|c|| c| }
		\hline  
		Neutral Higgs boson   &  ( Mass )${}^2$  \\  \hline
			 $H = \left( \upsilon \;h_a +  \widehat{\upsilon}\,{} \;\widehat{h}\,{}_a \right) / \upsilon_t $   
					&	$m^2_1$ \\  \hline
			 $H'	= \left( - \widehat{\upsilon} \;h_{a} + \upsilon \;\widehat{h}\,{}_a  \right) / \upsilon_t $ 
					& $m^2_2 $  \\[1ex]  \hline
					& \\[-1.25ex]
			$G^o	= \left( \upsilon \;z_a +  \widehat{\upsilon}\,{} \;\widehat{z}\,{}_a \right) / \upsilon_t $
					&  $m^2_Z $ \\[1ex]  \hline
					& \\[-1.25ex]
			$A	= \left( - \widehat{\upsilon} \;z_{a} + \upsilon \;\widehat{z}\,{}_a  \right) / \upsilon_t  $
					& $\lambda_1 \;\upsilon^2_t $  \\[1ex]  \hline
					& \\[-1.25ex]
			$\left. \begin{array}{l}
			    h_b, h_c, h_{ij} \;\;\;{\rm for} \;\;i \neq j  \\
					z_b, z_c, z_{ij} \;\;\;\;{\rm for} \;\;i \neq j 
							\end{array} \right\} $
					&  $\lambda_1 \;\widehat{\upsilon}\,{}^2 \;+\; \lambda_2 \;\upsilon^2 $ \\[1ex] \hline 
					&  \\[-1.5ex]
			$  \left. \begin{array}{l}
			    \widehat{h}_b, \widehat{h}_c, \widehat{h}_{ij} \;\;\;{\rm for} \;\;i \neq j  \\
					\widehat{z}_b, \widehat{z}_c, \widehat{z}_{ij} \;\;\;\;{\rm for} \;\;i \neq j 
							\end{array} \right\} $
					&   $ \lambda_1 \;\upsilon^2 \;+\; \lambda_2 \;\widehat{\upsilon}\,{}^2  $\\[1 ex] \hline
\end{tabular}
\end{center}
where, neglecting terms of order $\widehat{\upsilon}\,{}^4/\upsilon^2$, 
\begin{eqnarray}
		m^2_1 &=& ( 2 \lambda_2 - \lambda_3 ) \;\upsilon^2 + \displaystyle  \left[ \lambda_1
									+ \frac{ (\lambda_1 - \lambda_3 )^2}{2 \lambda_2 - \lambda_1 - \lambda_3 } \right] \;\widehat{\upsilon}\,{}^2 \\
		m^2_2 &=&   \lambda_1  \;\upsilon^2 \;+\;\; \displaystyle \left[ 2 \lambda_2 - \lambda_3
								-	\frac{ (\lambda_1 - \lambda_3 )^2}{2 \lambda_2 - \lambda_1 - \lambda_3 } \right] 
									\;\widehat{\upsilon}\,{}^2							
\end{eqnarray}
and we have, finally, the notation
\begin{eqnarray}
		\upsilon_t &\equiv& \sqrt{ \upsilon^2 + \widehat{\upsilon}\,{}^2} 
\end{eqnarray}

Note that the vacuum expectation values are
\begin{eqnarray}
		< h_{33}> 	&=& \sqrt{2} \,\upsilon_3   \nonumber \\
		< \widehat{h}_{33} >	&=&  \sqrt{2} \,\widehat{\upsilon}\,{}_3   \nonumber 
\end{eqnarray}
so that
\begin{eqnarray}
		m_t	&=& h_q \;\upsilon_3  \nonumber \\
		m_b	&=& h_q \;\widehat{\upsilon}\,{}_3 \nonumber
\end{eqnarray}

\section{Charged Higgs Boson Spectrum}

\begin{eqnarray}
		\left.  \begin{array}{lcl}
	\Phi^{\prime \,+}_a	&=& \displaystyle \left(\;\;\; \upsilon_3 \,\Phi^{+ \,'}_{33} \;+  \;\;\;\;\;\;
									\left( \upsilon_2 \,\Phi^{+ \,'}_{22}
								+ \upsilon_1 \,\Phi^{+ \,'}_{11} \right) \right) / \upsilon  \\
	\Phi^{\prime \,+}_b	&=& \displaystyle \left( -\upsilon_{\perp} \,\Phi^{+ \,'}_{33} 
							+ \frac{\upsilon_3}{\upsilon_{\perp}} \left( \upsilon_2 \,\Phi^{+ \,'}_{22} 
							 + \upsilon_1 \,\Phi^{+ \,'}_{11} \right) \right) / \upsilon   \\
	\Phi^{\prime \,+}_c	&=&  \;\;\;\;\;\;\;\;\;\;\;\;\;\;\;\;\;\;\;\;\;\;\;\;\; \left( - \upsilon_1 \,\Phi^{+ \,'}_{22} + \upsilon_2 \,\Phi^{+ \,'}_{11 } \right)
							/ \upsilon_{\perp}
						\end{array}  \right\}
\end{eqnarray}
with the notation
\begin{eqnarray}
	\Phi^{\prime \,+}_{ij}	&=& \Phi^{+}_{i \ell} \left( V^{\dagger} \right)^{\ell}_{j}
\end{eqnarray}
$\left( V \right)$ being the CKM matrix defined in eq.(\ref{CKM defined}). 

\begin{eqnarray}
		\left.  \begin{array}{lcl}
	\widehat{\Phi}\,{}^{\prime \,+}_{a}	&=& \displaystyle \left( \;\;\;\widehat{\upsilon}\,{}_3 \,\widehat{\Phi}\,{}^{' +33} 
				\;+  \;\;\;\;\;\;\left( \widehat{\upsilon}\,{}_2 \,\widehat{\Phi}\,{}^{'+22}
								+ \widehat{\upsilon}\,{}_1 \,\widehat{\Phi}\,{}^{' +11} \right) \right) / \widehat{\upsilon}\,{}  \\
	\widehat{\Phi}\,{}^{\prime \,+}_{b}	&=& \displaystyle \left( -\widehat{\upsilon}\,{}_{\perp} \,\widehat{\Phi}\,{}^{' +33} 
							+ \frac{\widehat{\upsilon}\,{}_3}{\widehat{\upsilon}\,{}_{\perp}} 
								\left( \widehat{\upsilon}\,{}_2 \,\widehat{\Phi}\,{}^{' +22}  
							 + \widehat{\upsilon}\,{}_1 \,\widehat{\Phi}\,{}^{' +11}  \right) \right) / \widehat{\upsilon}\,{}   \\
	\widehat{\Phi}\,{}^{\prime \,+}_{c}	&=&  \;\;\;\;\;\;\;\;\;\;\;\;\;\;\;\;\;\;\;\;\;\;\;\;\;\;\;\;\;  
								\left( - \widehat{\upsilon}\,{}_1 \,\widehat{\Phi}\,{}^{' +22} 
								+ \widehat{\upsilon}\,{}_2 \,\widehat{\Phi}\,{}^{' +11} \right) / \widehat{\upsilon}\,{}_{\perp}
						\end{array}  \right\}
\end{eqnarray}

\begin{eqnarray}
			\widehat{\Phi}\,{}^{\prime \,+ \,ij} &\equiv& \widehat{\Phi}\,{}^{+ \,i\ell} \,\left( V^{\dagger} \right)^j_{\ell}  
\end{eqnarray}

In terms of these basis states, we can now give the spectrum of charged Higgs mass eigenstates in the  't Hooft gauge ($\xi = 1$ )

\begin{center}
Table II \;\;\;
\begin{tabular}{|c|| c| }
		\hline  
		Charged Higgs boson   &  ( Mass )${}^2$  \\  \hline
		$G^+	= \left( \;\;\;\upsilon \;\Phi^{\prime \,+}_a +  \widehat{\upsilon}\,{} \;\widehat{\Phi}\,{}^{\prime \,+}_a \right) / \upsilon_t  $
					&	$ m^2_W $ \\  \hline
		$H^{\prime \,+}	= \left( - \widehat{\upsilon}\,{} \;\Phi^{\prime \,+}_a +  \upsilon  \;\widehat{\Phi}\,{}^{\prime \,+}_a \right) / \upsilon_t $
					& $\lambda_2 \;\upsilon^2_t $  \\[1ex]  \hline
					& \\[-1.25ex]
		$\Phi^{\prime \,+}_b \,, \Phi^{\prime \,+}_c \,, \Phi^{\prime \,+ \,ij} 
					\;for \;\;i \neq j $
					&  $ \lambda_1 \;\upsilon^2 \;+\; \lambda_2 \;\widehat{\upsilon}\,{}^2 $ 
					\\[1ex] \hline 
					&  \\[-1.5ex]
		$\widehat{\Phi}^{\prime \,+}_b \,, \widehat{\Phi}^{\prime \,+}_c \,, \widehat{\Phi}^{\prime \,+ \,ij} 
					\;for \;\;i \neq j  $
					&  $\lambda_1 \;\widehat{\upsilon}\,{}^2 \;+\; \lambda_2 \;\upsilon^2 $    
					\\[1 ex] \hline
\end{tabular}
\end{center}

\section{Higgs couplings to fermions}
Before considering the phenomenological implications of this Enriched Standard Model, we list here the couplings of the neutral Higgs mesons to the top and bottom fermions\footnote{\samepage
	For brevity, we have omitted here the Yukawa couplings involving the $u, d, c, s$ quarks.  Because of the hierarchy $\upsilon_3 >> \upsilon_2 >> \upsilon_1, \ldots$ they have negligible effect on the leading order phenomenology.
	}
\begin{eqnarray}
\left. 	\begin{array}{lcl}
	{\cal L}_{Y} &=& - \;\displaystyle \frac{h_q}{\sqrt{2}} \frac{\upsilon_3}{\upsilon}\;\bar{t} \;t \;\frac{\left( \upsilon \, H - \widehat{\upsilon}\, H' \right)}{\upsilon_t}   \;+\; i \;\frac{h_q}{\sqrt{2}} \frac{\upsilon_3}{\upsilon} \;\bar{t} \;\gamma_5 \;t \; \frac{ \left( \upsilon \;G^o - \widehat{\upsilon}\, A \right) }{\upsilon_{t}}  \\
		&&  -  \;\displaystyle \frac{h_q}{\sqrt{2}}\frac{\widehat{\upsilon}\,{}_3}{\widehat{\upsilon}} \;\bar{b} \;b \;\frac{\left( \widehat{\upsilon}\, H + \upsilon \, H' \right)}{\upsilon_t}   \;+\; i \;\frac{h_q}{\sqrt{2}}\frac{\widehat{\upsilon}\,{}_3}{\widehat{\upsilon}} \;\bar{b} \;\gamma_5 \;b \; \frac{ \left( \widehat{\upsilon}\, \;G^o + \upsilon \, A \right) }{\upsilon_{t}} 
				\end{array} \right\} 	\label{Higgs fermion coupling}
\end{eqnarray}
where $G^o$ is the Goldstone bosons in the 't Hooft gauge with mass $M_Z$.
Note that here $H$ and $H'$ are scalar fields, while $A$ is a pseudoscalar field.

\section{Phenomenological Implications}

There has been a lot of excitement in the world of particle physics ever since the announcement of the discovery of a $125 \;GeV$ Higgs-like boson on July 4, 2012, followed by the hint in late 2015 of a new resonance at $750 \;GeV$.  When I first proposed the ESM model in the context of the discovery of the $125 \;GeV$ Higgs, for simplicity of parameters, I restricted myself to the fully degenerate case of $\lambda_1 = \lambda_2 = \lambda_3$.  Now that there is the exciting possibility of a new $750 \;GeV$ Higgs, I present here a more general case of $\lambda_1 = \lambda_2$.  In a forthcoming paper, I will discuss the most general case, with all three different couplings, so that the Higgs family masses will have more fine structure.

\section{ $H$ field couplings same as those in SM}
Among the enriched family of neutral Higgs bosons, only the $H$ field develops a vacuum expectation value $\sqrt{2}\;\upsilon_t$.  Its trilinear and quartic couplings to the electroweak gauge fields and its Yukawa coupling to the matter fields are the same as in the Standard Model.  It can thus be identified with the SM Higgs field.

\section{ Only $H$ is in $W, Z$ fusion \& associated production}

While the VVH coupling is the usual SM coupling, eq.(\ref{covar deriv I}) leads to the complete trilinear decoupling of all the rest of the Higgs family.  By this I mean that $VVh_{ij}, \,VV\widehat{h}\,{}_{ij}, \,VVz_{ij}, 
\,VV\widehat{z}\,{}_{ij},   $ couplings for $(ij) \neq (33)$ are absent.  In the vector boson fusion \& associated production processes, therefore only $H$ is produced.  \\

\noindent The absence of VVH' coupling does not, however, imply a complete decoupling of $H'$ from the gauge fields. 
For they are coupled to the gauge bosons through the quartic coupling with the same coupling strengths as noted below
\begin{center}
		$\begin{array}{lclcl}
			{\cal L}_{VVHH} &=& {\cal L}_{VVH'H'} &=& {\cal L}_{VVH^{\prime \,+}H^{\prime \,-}} 
		\end{array}$
\end{center}
and likewise for the quartic coupling of the rest of the family of Higgs fields.

\section{Normal vs. Inverted Scenario}

Having determined the properties of $H$ versus $H'$, we now come to the interesting question of which of the two resonances is $H$, and which is $H'$.  For this discussion, we restrict ourselves to the case $\lambda_1 = \lambda_2$.  

\subsection{Normal Scenario}

If we identify the neutral $H$ state with the observed $125 \;GeV$, and the $H'$ state with the $750 \;GeV$, then the mass eigenvalues imply the values of the coupling constants

\begin{eqnarray}
		& & \left. \begin{array}{lcl}
		m_H &=& 125 \;GeV  \\
		m_{H'} &=& 750 \;GeV
		\end{array} 
		\right\} \begin{array}{rcr}
								\lambda_1, \lambda_2 &=&    18.59 \\
								\lambda_3	&=&		 36.66 
							\end{array}
		\label{lambda normal}
\end{eqnarray}

\noindent With this scenario, all the rest of the family of neutral as well as charged Higgs are crowded around $750 \;GeV$.  

\subsection{Inverted Scenario}

It is interesting to note the inverted scenario, where $H$ is the heavier resonance, while $H'$ is the lighter one.  In this case, we have
\begin{eqnarray}
		& & \left. \begin{array}{lcl}
		m_H &=& 750 \;GeV  \\
		m_{H'} &=& 125 \;GeV
		\end{array} 
		\right\} \begin{array}{rcr}
								\lambda_1, \lambda_2 &=&    0.516\\
								\lambda_3	&=&		-17.56 
							\end{array}
		\label{lambda inverted}
\end{eqnarray}
With this scenario, all the rest of the family of neutral as well as charged Higgs are crowded around $125 \;GeV$.  \\

\noindent In a forthcoming paper, we shall consider the case where $\lambda_1 \neq \lambda_2$, for which the rest of the family of Higgs will, in the normal scenario, be spread out above $750 \;GeV$, while in the inverted scenario, the rest of the family is now spread below $125 \;GeV$.

\section{ $H, H', A \rightarrow \gamma \gamma$ decay rates}

Preliminary data on the new $750 \;GeV$ resonance indicates a width for its $\gamma \gamma$ decay that is much wider than the $\gamma \gamma$ decay width for the $125 \;GeV$.  With the coupling constants as determined in eq.(\ref{lambda normal}, \ref{lambda inverted}), it is possible to make a leading order calculation of the predicted widths.  For this, we need to derive the trilinear Higgs self-coupling from $V_{\phi}$ in eq.(\ref{Higgs Pot})
\begin{eqnarray}
		{\cal L}_{HHH}	&=& - \displaystyle  \frac{ 2 \lambda_1 - \lambda_3 }{2}  \sqrt{2} \,\upsilon_t
										H \left( G^+ G_= + H'^+ \,H'_-      \right)  \nonumber \\
										& & - \frac{ 2 \lambda_1 - \lambda_3  }{2} \sqrt{2} \,\upsilon_t
										H \sum_{(ij) \neq (33)}\left(   H'^+_{ij} \,H^{\prime \,ij}_{-} 
											+ \widehat{H}\,{}^{\prime + \,ij} \,\widehat{H}\,{}^{\prime}_{- \,ij}      \right)
												\label{trilinear Higgs}
\end{eqnarray}

\noindent Based on eq.(\ref{Higgs fermion coupling}, \ref{trilinear Higgs}), we can proceed to make a leading order estimate of the decay rates in both the normal and inverted scenarios.  Using eq.(A4) of ref.\cite{Barroso}, the decay widths are

\begin{center}
Table III \;\;\;
\begin{tabular}{|c|| c| c| c|}
		 \hline  
		 & & & \\[.25ex]
		 Scenario &  $\Gamma( H \rightarrow \gamma \gamma)$ & $\Gamma( H' \rightarrow \gamma \gamma)$ & $\Gamma( A \rightarrow \gamma \gamma)$  \\[.25 ex]  \hline
		 & & & \\[-.15 ex]
		$m_H = 125 \;GeV, m_{H'} = 750 \;GeV$ 
					&	$ 10 \;keV$  & $0.7 \;keV$ & $0.5 \;keV$ \\[.25 ex]  \hline
					& & &\\[-.15ex]
		$m_H = 750 \;GeV, m_{H'} = 125 \;GeV$ 
					&	$ 45.6 \;MeV$  & $1.4 \;keV$ & $1.3 \;keV$ \\[.25 ex]  \hline
\end{tabular}
\end{center}

\noindent What is interesting about the Inverted Scenario is the large width for $H \rightarrow \gamma \gamma$ compared with the $2 \gamma$ decay width of the smaller mass $H'$ Higgs.

\section{ Conclusion}
\noindent There has been a lot of excitement in the world of particle physics ever since the announcement of the discovery of Higgs boson on July 4, 2012, and, over the Christmas break of 2015, the hint of a new resonance at $750 \;GeV$.  Now that we contemplate an Enriched Standard Model, a natural question that arises would be which of the two is the $H$ and $H'$ of our family of neutral Higgs.  The answer becomes clear when we take into account the production and decay channels involved. \\

\noindent For at high energies, the leading production processes predominantly involve top quark loops.  Therefore, the production of the family Higgs, $h_{ij}$, $z_{ij}$, $\widehat{h}\,{}_{ij}$, $\widehat{z}\,{}_{ij}$, $\Phi^{\prime \,+}_{ij}, \;\widehat{\Phi}\,{}^{\prime \,+ \,ij}$, with $(ij) \neq (33)$, are suppressed. \\

\noindent Among the family of Higgs, $H$, plays a special role.  It behaves like the single Higgs field of the Standard Model, with the same trilinear coupling to the gauge bosons as the standard Higgs.  It is produced via associative production with $W, Z$ or Vector Boson fusion.  Its production cross-section is identical to that of the Standard Model.  \\

\noindent In contrast, the orthogonal Higgs bosons, $H'$ and $A$, do not have trilinear couplings to the gauge bosons.  They are produced through coupling to the top and bottom quarks, see eq.(\ref{Higgs fermion coupling}).  While we have not yet pursued a full-blown program of calculating all the widths, the partial result involving the $2 \gamma$ decay encourages us to look for dramatic difference in the widths.

\noindent There is much work that remains to be done to explore the consequences of this proposal.  I welcome your comments and suggestions.  \\

\section{Acknowledgments}
\noindent I wish to thank IAS and Nanyang Technological University for the warm hospitality over the consecutive summers where this work was done.  I also wish to thank and acknowledge Zhi-Zhong Xing, Markos Maniatis, and Manmohan Gupta for their many stimulating discussions at IAS on quark texture matrices, and Higgs phenomenology.

\end{document}